\documentclass{ws-p8-50x6-00}
\def\be{\begin{equation}}
\def\ee{\end{equation}}
\def\bea{\begin{eqnarray}}
\def\eea{\end{eqnarray}}

\begin{document}

\title{Modelling correlations in heavy ion collisions}

\author{T. Cs\"org\H{o}}

\address{$^1$ Dept. Phys. Columbia University,\\
	 538 W 120th St New York, NY -10027, USA\\
	$^2$ MTA KFKI RMKI, H - 1525 Budapest XII,
	Konkoly-Thege 29-33, Hungary, \\
	E-mail: csorgo@sunserv.kfki.hu}

\author{A. Ster}

\address{MTA KFKI RMKI and MFA,\\
	 H - 1525 Budapest XII, Konkoly-Thege 29-33, Hungary, \\
	E-mail: ster@rmki.kfki.hu}

\maketitle

\abstracts{
Modelling correlations in heavy ion collisions is reviewed
from model independent characterization of 
the two-particle correlations to  
 Buda-Lund hydro model fits to  correlations and spectra 
in Au + Au collisions at $\sqrt{s}=130$ AGeV at RHIC.
}

\section{Recent results in modelling correlations}
Modelling correlations in heavy ion reactions is 
a broad and rapidly expanding field that 
is reviewed here very briefly, due to space limitations.
For a more detailed and illustrated
version of this talk, see ref.\cite{cs-ismd01}. 
For more detailed recent reviews see also refs.\cite{csrev,krev}.

Although some claims were made in the literature that
the shape of the two-particle Bose-Einstein correlation 
functions has to be a (multi-variate) Gaussian, by now it is
clear\cite{sk,hardtke} that these claims were mistaken and even
approximately Gaussian correlation functions may contain subleading
non-Gaussian corrections, for example oscillatory prefactors\cite{csrev}
that are particularly important in case of effective 
binary sources\cite{binary}.  It is also clear that refined analysis of the
two-particle correlation functions indicates significant deviations
from Gaussian shapes in elementary particle induced 
reactions\cite{krev}.
Sensitive searches for non-Gaussian components have not yet been
performed in correlation measurements 
in heavy ion collisions at CERN SPS and at RHIC, to our best knowledge. 
A model independent parameterization of the 
two-particle correlation 
function is given in terms of the Edgeworth or Laguerre expansions\cite{edge}.
These expansions were shown\cite{csrev,krev,edge} 
to be sensitive to even small deviations from
Gaussian or exponential shapes.

The deviation from an exactly Gaussian form of the two-particle
correlation function can be caused by the decay products of 
long-lived resonances. A core-halo picture has been introduced
 to describe such a situation\cite{sk,chalo,halo2,halo3}.
Within the core-halo picture, the intercept parameter 
$\lambda_*$ of the
two-particle correlation function is interpreted as the
squared, momentum dependent fraction  of particles emitted from
the core of the interaction, and the shape of the correlation
function is interpreted as a measure of the core part of the 
interaction region.
The core-halo structure of the particle emitting source
has recently been observed with the help of a new imaging
method, developed by D. A. Brown and P. Danielewitz\cite{image}, 
that reconstructs the 
relative source distribution directly from the measured data points
with the help of the known final state interactions of the emitted pair.

Good progress has been made recently
in the study of multi-particle Coulomb corrections. 
A wave-function integration method has been developed for
3 - 5 particles, utilizing a cluster decomposition of the multiparticle
final state\cite{c3,cn}. 
Such Coulomb corrections have to be applied within the context 
of a core-halo picture\cite{biya,alt-tihany}.
A modification of the Coulomb interactions (permutations in
the charge allocation) has recently been proposed to develop 
Monte-Carlo event generators with Bose-Einstein correlations\cite{utyuzh}.

An interesting new direction of modelling correlations in heavy ion
collisions is the study of back-to-back correlations (BBC)
for bosonic\cite{ac,acg,acg2,aw}  as well as
fermionic cases\cite{fbbc} (bBBC and fBBC). 
When the thermalized vacuum state decays, 
it always produces particle - anti-particle pairs with opposite
spin and momenta. Hence the strength of BBC is increasing inversely with
the single particle spectra and in principle, it can be unlimitedly 
large\cite{acg,fbbc}. It would be certainly of great interest
to observe BBC experimentally.  The bosonic 
BBC in heavy ion physics has a famous analogy in
astropysics, namely  the Bekenstein-Hawking radiation
of black holes\cite{Bekenstein:1973ur,Hawking:1975sw}. 
Both effects are related to squeezing and decay of a
modified vacuum to pairs of  bosons of asymptotic fields
with back-to-back momentum and opposite quantum numbers.

\section{Hydro models fit spectra at RHIC}
Modelling high energy heavy ion collisions is not only difficult because
of the large number of degrees of freedom involved in the process but
also because various theoretical concepts and approaches are relevant for
different stages of the collision. A comprehensive review of 
this process has recently been given by 
S.A. Bass\cite{bassc}.
During a partonic cascading process
(local) thermalization is achieved, 
that can be maintained for a while due to the intensive 
collision rate and the time evolution corresponds to 
nuclear fluid dynamics.
As the system rarifies, the collision rate within the ``macroscopically 
infinitesimal" fluid cells decreases and one expects that the hydrodynamical
approximation may break down and non-equilibrium hadronic
transport models start to play a role\cite{bassc}. 
A version of this well established concept has been utilized to
describe the single particle spectra and two-particle correlations
in 130 AGeV Au + Au collisions at  RHIC\cite{soffbd}.
Numerical solution of relativistic hydrodynamics was terminated by
particle freeze-out and a subsequent resonance decay and hadronic cascading
by a code developed by Soff, Bass and Dumitru.
The model fitted the single particle spectra measured by the PHENIX and STAR 
experiments at RHIC\cite{PHENIXimd-qm01,STARimd-qm01},
but it overestimated, in a statistically unacceptable manner, the experimentally
observed longitudinal and out radius components of the Bose-Einstein
correlation functions\cite{PHENIXhbt-qm01,STARhbt-qm01}. 
Both the longitudinal and the out radius components are sensitive to
the distribution of particle production in time. 
One of the problems with the calculation could be its utilization of the 
Gaussian variances\cite{uli_s,uli_l,urev} 
of the source distribution, that are known to 
overestimate the HBT radii for core-halo type of 
systems\cite{csrev,sk,hardtke,3d}.

An interesting study of the identified particle elliptic flow at RHIC has
been reported by  the STAR collaboration\cite{snellings}. From the mass
dependence of the second flow coefficient as a function of transverse
momentum, a suggestive evidence has been obtained that early local
equilibration followed by a hydrodynamic expansion is responsible for
the particle spectra in central and mid-peripheral collisions up to
$p_t \le 1.5$ GeV. A similar conclusion has been reached by Kolb,
Huovinen, Heinz and Heiselberg, who observed in addition that the
very rapid thermalization in Au + Au collisions at RHIC provides a
serious challenge for kinetic approaches based on classical
scattering of on-shell particles\cite{ellheinz}.
On the other hand, the high $p_t$ behaviour of the elliptic flow pattern
was shown to be a sensitive measure of the initial parton density distribution
of an initial quark-gluon plasma phase\cite{ellgyu}.
Detailed numerical solution of relativistic hydrodynamics
has been reported to describe well the single particle spectra,
the first and the second flow coefficients for various domains of
centrality and transverse momentum\cite{teaney}, 
however, this calculation also over-predicted  by a factor of 2
the measured, effective HBT radii of 
the two-particle Bose-Einstein correlation function, 
possibly indicating a 
problem with the freeze-out time distribution in this kind of calculations.
A good description of various aspects of  the measured single particle
spectra, and the $m_t$ dependence of the longitudinal HBT radius parameters
has been reported by Hirano, Morita, Muroya and Nonaka\cite{jhyd}
 based on an exact solution of relativistic hydrodynamics.
The order of magnitute of the calculated side and out radii were also
found to be in agreement with the experimental data, however,
the side radius component was slightly underestimated and the out
radius component over-estimated by this model, and the observed
decrease of the out component with increasing transverse mass
has not been reproduced.

\section{Buda-Lund hydro fits spectra and HBT radii at RHIC }
It has been observed that simple parameterizations of the freeze-out
phase space distribution of 
a locally thermalized, three dimensionally expanding, cylindrically
symmetric finite systems describe the preliminary single particle
spectra at RHIC reasonably well\cite{florkowski} in the whole available
kinematic domain.

The Buda-Lund hydrodynamical parameterization (BL-H) has been developed 
in refs.\cite{3d,qm95} to describe particle correlations and
spectra in heavy ion collisions at CERN SPS energies. The model
was formulated on the principle of cylindrical symmetry in central
collisions and a relativistic, 3 dimensional flow profile that corresponds to
a Hubble type of transverse flow sitting on top of a longitudinal
Bjorken flow.  Under certain conditions\cite{3d}, the BL-H parameterization
predicted an $R_l \simeq R_s \simeq R_o \propto 1/\sqrt{m_t}$
scaling of the HBT radius parameters, as a direct reflection of the
cylindrical symmetry of the source. 
\begin{figure}[t] 
\null\vspace*{-0.5cm}
%\figurebox{20pc}{15pc}{} % to have a box alone 
\epsfxsize=14pc % will enlarge or reduce based on the xsize 
\epsfbox{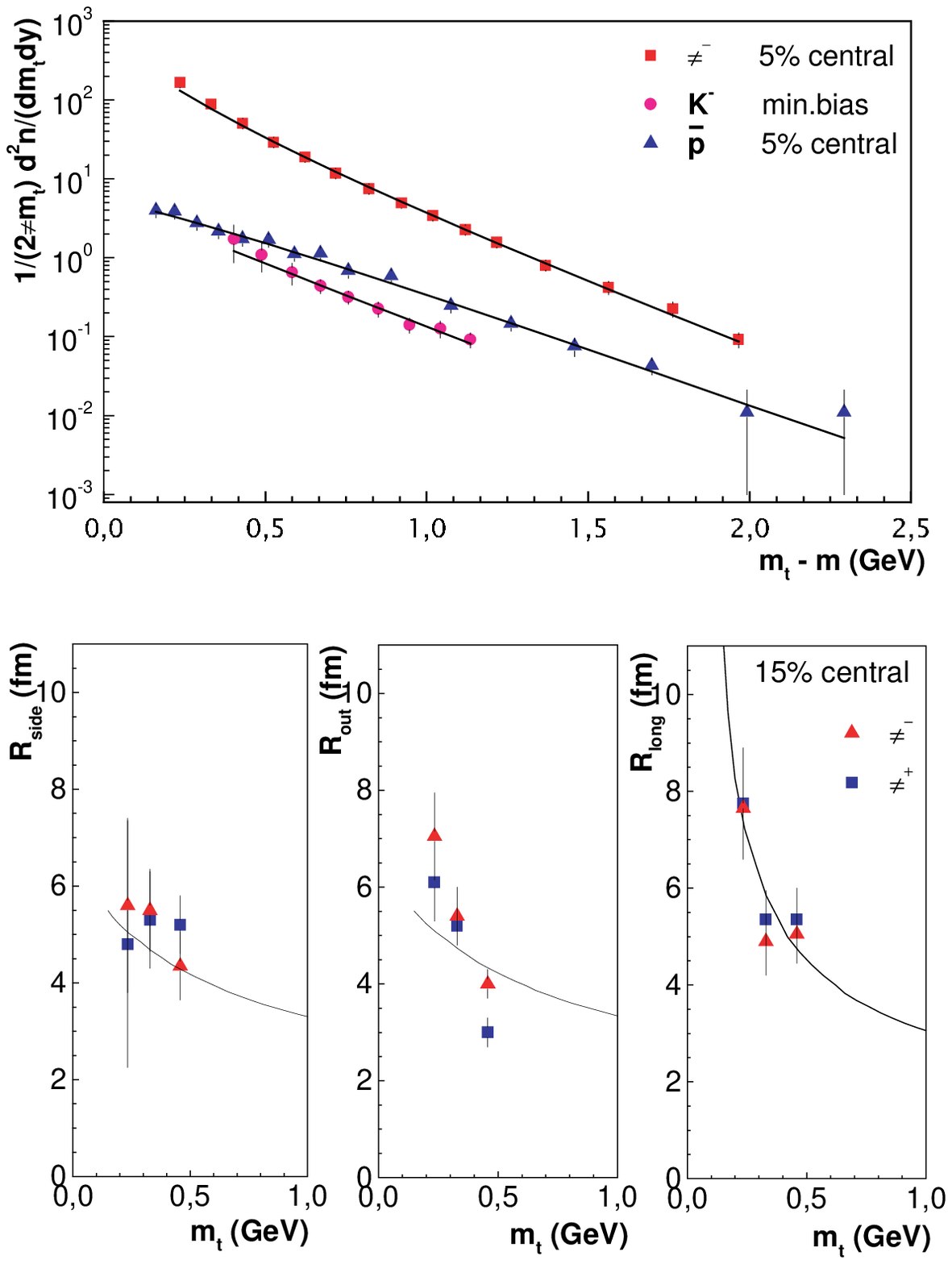} % postscript image file name 
\epsfxsize=15pc % will enlarge or reduce based on the xsize 
\epsfbox{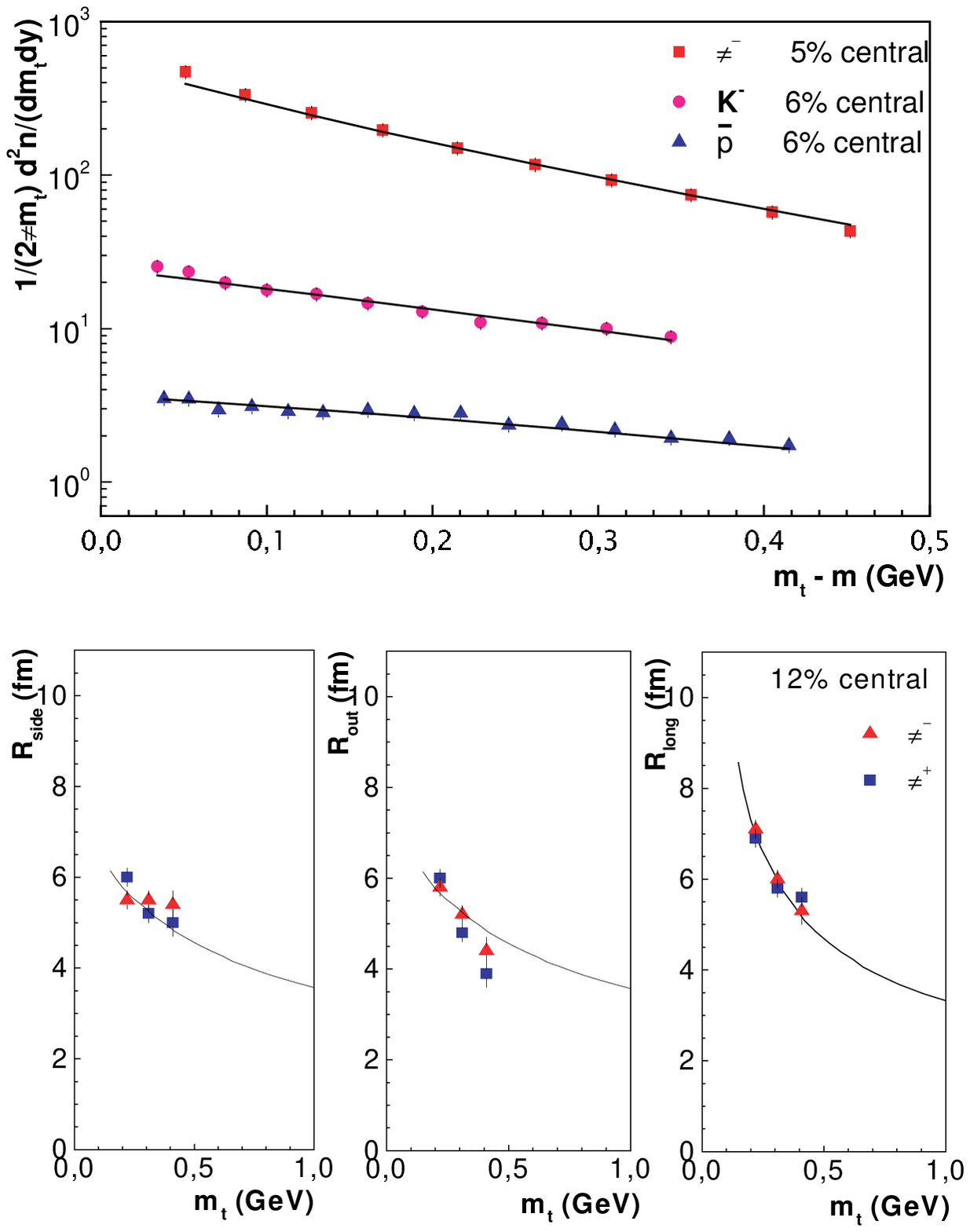} % postscript image file name 
\vspace*{-1.0cm}
\caption{
\label{fig:Phenix-BL}
Simultaneous fit of the Buda-Lund hydro model
to preliminary data of PHENIX$^{40,42}$
(left panels), and 
STAR$^{41,43}$
(right panels) 
on particle spectra and correlations of
central Au + Au collisions at RHIC, $\sqrt{s} = 130 $AGeV.
} 
\end{figure}          

\begin{table}[tb]
% -----------------------------------------------------
% adapted from TeX book, p. 241
\newlength{\digitwidth} \settowidth{\digitwidth}{\rm 0}
\catcode`?=\active \def?{\kern\digitwidth}
% -----------------------------------------------------
\caption{Source parameters from simultaneous fittings of 
 CERN SPS (NA49, NA44 and preliminary WA98) and RHIC (preliminary
 PHENIX and STAR) particle spectra and HBT radius parameters with
 the Buda-Lund hydrodynamical model.}
\label{tab:results}
\begin{center}
\footnotesize
\begin{tabular*}{\textwidth}{@{}|l@{\extracolsep{\fill}}|rl|rl||rl|rl|rl|}
\hline
                 BL-H
                 & \multicolumn{2}{l|}{STAR prel.}
                 & \multicolumn{2}{l||}{PHENIX prel.} 
                 & \multicolumn{2}{l|}{$\langle RHIC \rangle$} 
                 & \multicolumn{2}{l|}{$\langle SPS \rangle$} 
                 \\
\cline{2-9}  
                 parameters~\,~\,
                 & \multicolumn{1}{r}{Value} 
                 & \multicolumn{1}{l|}{Error} 
                 & \multicolumn{1}{r}{Value} 
                 & \multicolumn{1}{l||}{Error} 
                 & \multicolumn{1}{r}{Value} 
                 & \multicolumn{1}{l|}{Error} 
                 & \multicolumn{1}{r}{Value} 
                 & \multicolumn{1}{l|}{Error} 
                 \\
\hline
$T_0$ [MeV]      & 143  &$\pm$ 4    & 140  &$\pm$ 3     & 142  &$\pm$ 2    & 139  &$\pm$ 6 \\
$\langle u_t \rangle$
                 & 0.76 &$\pm$ 0.06 & 0.67 &$\pm$ 0.3   & 0.71 &$\pm$ 0.05 & 0.55 &$\pm$ 0.06\\
$R_G$ [fm]       & 7.5  &$\pm$ 0.3  & 6.6  &$\pm$ 0.3   & 7.1  &$\pm$ 0.7  & 7.1  &$\pm$ 0.2\\
$\tau_0$ [fm/c]  & 8.8  &$\pm$ 0.5  & 7.8  &$\pm$ 0.3   & 8.3  &$\pm$ 0.7  & 5.9  &$\pm$ 0.6\\
$\Delta\tau$ [fm/c]
                 & 0.01 &$\pm$ 1.2  & 1.3  &$\pm$ 1.0   & 1.3  &$\pm$ 1.0  & 1.6  &$\pm$ 1.5\\
$\Delta\eta$     & 1.0  &$\pm$ 0.9  & 1.5  &$\pm$ 0.1   & 1.3  &$\pm$ 0.4  & 2.1  &$\pm$ 0.4\\
$\langle {\Delta T \over T}\rangle_r$
                 & 0.09 &$\pm$ 0.02 & 0.01 &$\pm$ 0.02  & 0.05 &$\pm$ 0.05 & 0.06 &$\pm$ 0.05\\
$\langle {\Delta T \over T}\rangle_t$ 
                 & 1.4  &$\pm$ 0.4  & 0.9  &$\pm$ 0.1   & 1.2  &$\pm$ 0.4  & 0.59 &$\pm$ 0.38\\
\hline
$\chi^2/NDF$     & \multicolumn{2}{l|}{46/58 = 0.79} 
                 & \multicolumn{2}{l||}{45/54 = 0.83} 
                 & \multicolumn{1}{r}{0.81} 
                 & \multicolumn{1}{r|}{ } 
                 & \multicolumn{1}{r}{1.20} 
                 & \multicolumn{1}{l|}{ } 
                 \\
\hline
\end{tabular*}
\end{center}
\end{table}
Both the single particle spectra and the two-particle Bose-Einstein
correlation functions are analytically calculated
\cite{csrev,3d,qm99,ster-hip99} as a function of 8 model parameters plus
constants of normalization for the absolutely normalized single particle
spectra.  The BL-H model fits simultaneously the double differential 
single particle spectra and the $m_t$ dependence of the 
radius parameters in hadron-proton reactions at CERN SPS\cite{na22}.
The BL-H also  describes the NA49, NA44
and the preliminary WA98 data for spectra and correlations 
at CERN SPS, with a
statistically  acceptable $ \chi^2/NDF$ fit\cite{qm99},
 as summarized in the last column of
Table 1 for a comparision with 
BL-H fits to the preliminary PHENIX and STAR data for
Au + Au at $\sqrt{s} = 130$ AGeV at RHIC. We observe that the 
BL-H fits of these RHIC data are satisfactory. The central value
of the freeze-out temperature, $T_0$ is about the same at RHIC as
at CERN SPS, and within three standard deviations, most  of the fit
parameter values are similar. However, 
the mean value of the freeze-out time $\tau_0$, 
and the strength of the  transverse flow at the geometrical
radius, $\langle u_t \rangle$, is significantly
larger at RHIC than at CERN SPS. The system at RHIC freezes
out later, with a larger transverse flow parameter 
$\langle u_t \rangle$ than at CERN SPS. 
The preliminary results indicate that within
the presently big errors the transverse geometrical size 
$R_G$ is about as big at RHIC as at CERN SPS.
This research was supported by grants
OTKA T026435 T034296 N25487 (with NWO), NSF-MTA-OTKA 0089462 and US DOE 
DE - FG02 - 93ER40764.

\end{document}